\documentclass[a4paper,10pt,aps,prb,twocolumn,showpacs,floatfix]{revtex4}

\usepackage{amssymb}
\usepackage{amsmath}
\usepackage{graphicx}

\usepackage{color}

\providecommand{\cite}[1]{\textcolor{blue}{\cite{#1}}}
\providecommand{\footnote}[1]{\textcolor{blue}{\footnote{#1}}}

\let\oldsqrt\sqrt
\def\sqrt{\mathpalette\DHLhksqrt}
\def\DHLhksqrt#1#2{%
\setbox0=\hbox{$#1\oldsqrt{#2\,}$}\dimen0=\ht0
\advance\dimen0-0.2\ht0
\setbox2=\hbox{\vrule height\ht0 depth -\dimen0}%
{\box0\lower0.4pt\box2}}

\usepackage[colorlinks]{hyperref}

\begin{document}
\title{Low Bias Negative Differential Resistance in Graphene Nanoribbon Superlattices}

%
%
%

\author{Gerson J.~Ferreira,$^{1,2}$ Michael N.~Leuenberger,$^{2}$ Daniel Loss,$^{3}$ J.~Carlos Egues$^{1,3}$}
\affiliation{
$^1$Departamento de F\'{\i}sica e Inform\'{a}tica, Instituto de F\'{\i}sica de S\~{a}o Carlos, Universidade de S\~{a}o Paulo, 13560-970 S\~{a}o Carlos, S\~{a}o Paulo, Brazil\\
$^2$NanoScience Technology Center and Deptartment of Physics, University of Central Florida, 12424 Research Parkway Suite 400, Orlando, Florida 32826, USA\\
$^3$Department of Physics, University of Basel, Klingelbergstrasse 82, CH-4056 Basel, Switzerland
}

\date{\today}

\begin{abstract}
We theoretically investigate negative differential resistance (NDR) for ballistic transport in semiconducting armchair graphene nanoribbon (aGNR) superlattices (5 to 20 barriers) at low bias voltages $V_\text{SD} < 500$~mV. We combine the graphene Dirac Hamiltonian with the Landauer-B\"uttiker formalism to calculate the current $I_\text{SD}$ through the system. We find three distinct transport regimes in which NDR occurs: (i) a ``classical'' regime for wide layers, through which the transport across band gaps is strongly suppressed, leading to alternating regions of nearly unity and zero transmission probabilities as a function of $V_\text{SD}$ due to crossing of band gaps from different layers; (ii) a quantum regime dominated by superlattice miniband conduction, with current suppression arising from the misalignment of miniband states with increasing $V_\text{SD}$; and (iii) a Wannier-Stark ladder regime with current peaks occurring at the crossings of Wannier-Stark rungs from distinct ladders. We observe NDR at voltage biases as low as $10$~mV with a high current density, making the aGNR superlattices attractive for device applications.
\end{abstract}
\pacs{72.80.Vp, 73.22.Pr, 73.21.Cd, 68.65.Cd}
\maketitle

\section{Introduction}
Graphene \cite{novoselov2005two,KatsnelsonNature2006,CastroNeto2009} has attracted much attention due
to the possibility of new devices that may surpass their semiconductor
counterparts in both speed and reduced power consumption.
\cite{avouris2010graphene} This is expected due to the unique properties of
graphene, e.g., the high mobility of carriers, which can lead to high current
densities, and the tunability of the bandgap. Additionally, building devices on
the surface could facilitate optical absorption and emission. Particularly,
negative differential resistance (NDR) is essential for many applications.
\cite{Cardona_book,Mortazawi,Sollner,sze2007physics} In semiconductor resonant
tunneling diodes \cite{Tsu1973,SollnerAPL1983,Iogansen1964} and superlattice
structures, \cite{Esaki_Tsu,Tsu_book} NDR is based on Fabry-P\'erot-type
interferences arising from the impedance mismatch between the various layers.
These semiconductor NDR systems can also show interesting phenomena, such as
intrinsic bistability due to charge accumulation. \cite{Goldman1987} Pursuing
the recent interest in graphene superlattices transport and thermal properties,
\cite{bai2007klein,brey2009emerging,park2009landau,
barbier2010extra,stojanovic2010polaronic, burset2011transport,
guo2011conductance,jiang2011minimum,abedpour2009conductance,cheraghchi2011metallic} it is a natural question to ask whether a
graphene superlattice could exhibit similar features.

\begin{figure}[ht!]
  \includegraphics[width=8cm]{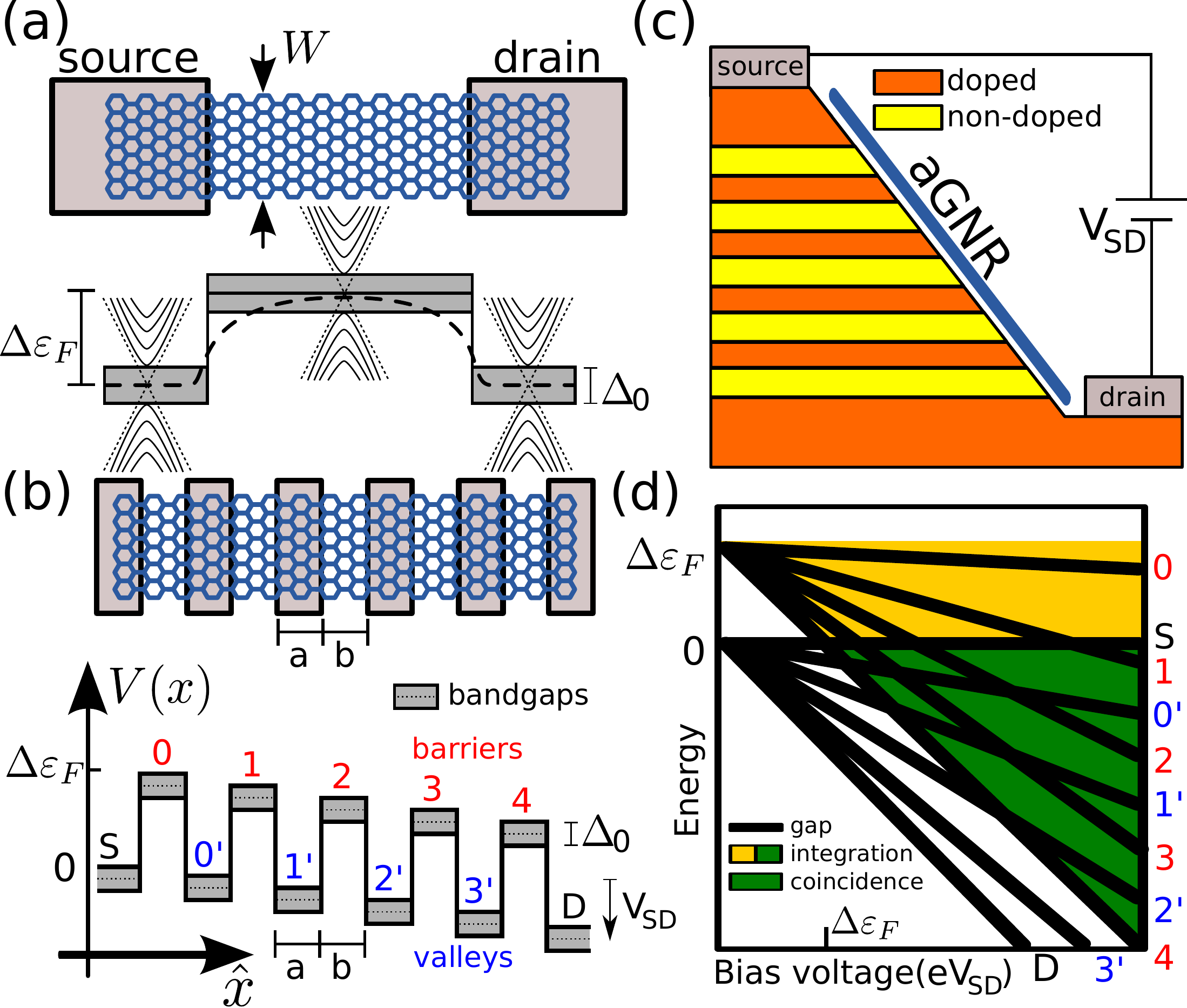}
  \caption{(Color online) (a) Metal-aGNR junctions and the modulated chemical shift $\Delta \varepsilon_\text{F}$ of the Dirac point across the aGNR  (Refs.~\onlinecite{GiovannettiPRL2008,VaninPRB2010,SalvadorPRL2010,Varykhalov2010}). $\Delta_0$ (shaded regions) denotes the barrier and valley bandgaps. Here we consider square potentials, solid line. The dashed line shows the numerical results of Ref.~\onlinecite{SalvadorPRL2010}. (b) Additional electrodes modulate the Dirac cone shift into a superlattice potential. The bias voltage $V_\text{SD}$ is also shown. (c) Doped layers of a semiconductor superlattice can also modulate the local potential. (d) Schematic of the $\varepsilon-V_\text{SD}$ diagram of the source--drain transmission coefficient showing crossings of the bandgaps $\Delta_0$ (black lines). The shaded regions delimit the energy range between the source $\mu_\text{S} = \Delta\varepsilon_\text{F}$ and drain $\mu_\text{D} = \mu_\text{S}-eV_\text{SD}$ chemical potentials.}
  \label{fig1}
\end{figure}

The occurrence of Klein tunneling in graphene \cite{KatsnelsonNature2006} should be an obstacle to the NDR effect, as it gives a monotonically increasing contribution to the current. Narrow graphene nanoribbons overcome this limitation as the lateral confinement quantizes the Dirac cone into few-eV-wide bands. Tight-binding calculations show that it is possible to find NDR in these narrow nanoribbons at high bias voltages, $1$--$2$~V. \cite{WangAPL2008,NamDoJAP2010} However, for integrated circuits a low bias mV regime is desirable to reduce power consumption.
\cite{footnoteDragoman}
Low bias NDR can also be achieved in other graphene and bilayer graphene systems. \cite{RenAPL2009,habib2011negative,fang2011strain}

\begin{figure}
  \includegraphics[width=8cm]{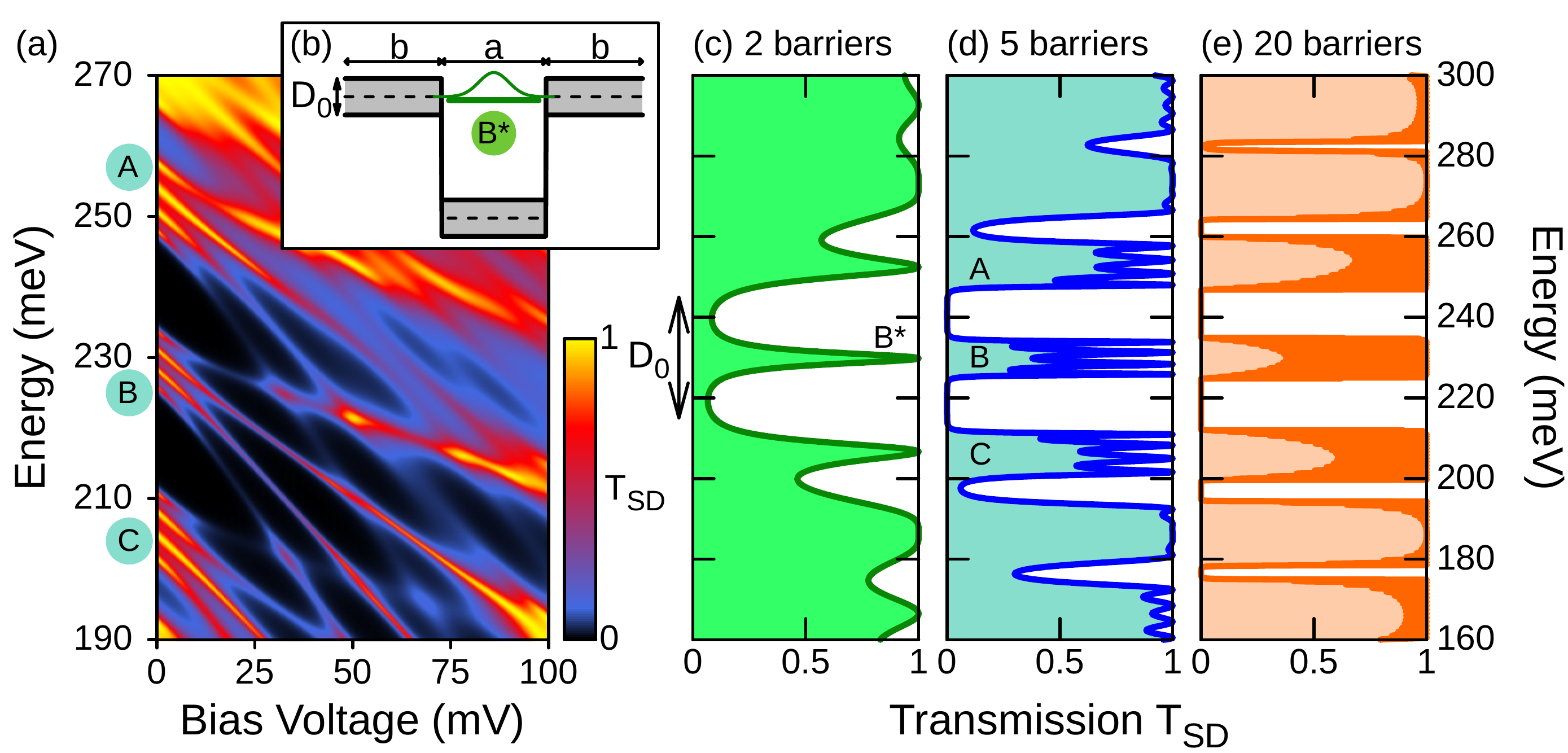}
  \caption{(Color online) (a) Energy-voltage diagram of $T_\text{SD}$ for $N=5$ barriers showing the evolution of the $N-1$ hybridized modes [panels (b)--(d)] into Wannier-Stark ladders. Labels A, B, and C show the zero-bias hybridized modes in panels (a) and (d). Crossings of ladders' rungs from distinct minibands increases $T_\text{SD}$ near $V_\text{SD}=30$ and $50$~mV. (b) Schematic of the modulated Dirac point (dashed line), bandgaps $\Delta_0 \sim 28$~meV (gray area), and confined mode $B^\prime$. In the transmission coefficient $T_\text{SD}$ across two barriers ($a=b=50$~nm) (c), the confined mode $B^\prime$ shows up as a resonant spike near $230$~meV. For (d) $N=5$, and (e) $N=20$ barriers the confined modes hybridize into $N-1$ spikes, building up a miniband. Similar resonances lead to minibands at energies away from the bandgap region $\Delta_0$.}
  \label{fig2}
\end{figure}

In this work we consider an $N$-barrier superlattice potential on a semiconducting armchair graphene nanoribbon (aGNR); Fig.~\ref{fig1}. The electronic structure of the aGNR is a quantized Dirac cone, due to the quantization of the transversal momentum $k_n$, and can be metallic, $k_{n_0} = 0$, or semiconducting, $k_{n_0} \neq 0$, depending on the width $W$ of the nanoribbon; $k_{n_0}$ is the closest to zero transverse momenta. We choose $W = 346a_0$, such that the aGNR is semiconducting with a bandgap $\Delta_0 = 28$~meV; $a_0 = 0.142$~nm is the C-C distance. We use the transfer-matrix formalism to calculate the source-drain transmission coefficient $T_\text{SD}$ across the superlattice potential along the aGNR, considering a finite bias voltage $V_\text{SD}$, revealing the electronic structure of the system; Fig.~\ref{fig2}.
The potential drop from source to drain follows a piecewise constant profile layer by layer; Fig.~\ref{fig1}(b).
The current is calculated within the usual Landauer-B\"uttiker formalism.

We find low bias NDR at zero and room temperatures within three distinct
physical regimes. (i) For wide layers, the transmission across the bandgaps
$\Delta_0$ is strongly suppressed, and nearly unity for energies away from the
bandgaps. With increasing voltage, both barrier and valley bandgaps split and
cross as shown schematically in Fig.~\ref{fig1}(d), showing, at the coincidence
region, a pattern of diamond-shaped structures with alternating regions of
finite and suppressed transmission, thus leading to NDR.
For narrow barriers resonant tunneling across layers become relevant. (ii) At
zero bias, hybridization of resonant modes leads to minibands with finite,
nearly unity, transmission; Fig.~\ref{fig2}(b)-\ref{fig2}(e). At very low voltages
$eV_\text{SD} \sim 10$~meV (of the order of the miniband energy width) the resonant
states misalign, thus breaking the minibands into off-resonance Wannier-Stark
ladders with suppressed transmission. This gives rise to a single current spike
near $eV_\text{SD} \sim 10$~meV. (iii) With increasing $eV_\text{SD}$, rungs of ladders
from distinct minibands cross and hybridize, showing a new set of resonant
spikes in $T_\text{SD}$, Fig.~\ref{fig2}(a), thus leading to current spikes and NDR.

\section{Proposed system \& model} The modulation of the Dirac cone into a superlattice
potential can be achieved by different setups. It was shown that local charge-transfer effects between graphene and some metals (e.g., Al, Cu, Ag, Au, Pt)
rigidly shifts the Dirac cone;
\cite{GiovannettiPRL2008,VaninPRB2010,SalvadorPRL2010,Varykhalov2010}
Fig.~\ref{fig1}(a). A series of metallic stripes over graphene can create the
proposed superlattice potential; Fig.~\ref{fig1}(b). Equivalently, the same
structure can be obtained by selectively doping graphene regions in an alternate
fashion. Additionally, the aGNR could be arranged along the doped/non-doped
layers of a cleaved semiconductor heterostructure; \cite{Krahne2002}
Fig.~\ref{fig1}(c). Narrow systems ($\lesssim 400$~nm) are desirable to keep
transport ballistic at room temperatures.

We consider low-energy excitations of graphene
within the envelope function approximation, \cite{Wallace1947,CastroNeto2009}
i.e., the graphene Dirac Hamiltonian. The finite size of the nanoribbon requires
vanishing wave functions at the edges, where for aGNR both $A$ and $B$ sublattices
of the honeycomb lattice are present. This leads to vanishing boundary
conditions for the envelope functions at these edges. \cite{CastroNeto2009} The
validity of these boundary conditions is discussed in
Ref.~\onlinecite{BreyFertigPRB2006}. Within this description, the electronic
structure of an aGNR is a quantized Dirac cone, $\varepsilon = s \hbar
v_\text{f}\sqrt{k_x^2+k_n^2}$. Here $s = \pm 1$ for the conduction and valence bands,
$v_\text{f} \approx 10^{15}$~nm/s is the Fermi velocity, $k_x$ is the momentum in the
longitudinal direction $\hat{x}$, $k_n = n\pi/W - 4\pi/3a_0$ is the quantized
transverse momentum with integer $n$, and $W = 346a_0 \sim 50$~nm. The
fundamental gap is given by $\Delta_0 = 2\hbar v_\text{f} |k_{n_0}| = 28$~meV, with
$k_{n_0} \sim -0.021$~nm$^{-1}$.

To calculate the transmission
$T_\text{SD} \equiv T_\text{SD}(\varepsilon, k_n, V_\text{SD})$ we use the transfer-matrix
formalism, \cite{Datta95} which relates the coefficients of the incoming and outgoing plane waves at the
source and drain leads across the superlattice layers (see the Appendix for details). We consider a piecewise
constant superlattice potential along the $x$ direction, Figs.~\ref{fig1}(b),
through which the electronic structure of each layer is shifted by the local
potential. In Figs.~\ref{fig2}--\ref{fig4} we show $T_\text{SD}$ only for $k_{n_0}$,
as it contains the major contribution for the current in all investigated cases.

The current density of Dirac electrons in
graphene is given by $\vec{\jmath}(\mathbf{r}) = 4ev_\text{f}
\psi^\dagger(\mathbf{r})\vec{\sigma}\psi(\mathbf{r})$, where the factor of $4$
accounts for the valley and spin degeneracies, $\psi(\mathbf{r})$ is the
envelope function spinor for the $K$ or $K^\prime$ valley, and $\vec{\sigma} =
(\sigma_x,\sigma_y)$ are the Pauli matrices. Within the Landauer-B\"uttiker
formalism, \cite{Datta95,BlanterButtiker2000} the current reads

\begin{equation}
 I_\text{SD} = \dfrac{e}{h} \sum_n \int_{-\infty}^{\infty} T_\text{SD}(\varepsilon, k_n,
V_\text{SD})\left[f_\text{S}(\varepsilon)-f_\text{D}(\varepsilon)\right] d\varepsilon,
\end{equation}
where $f_\text{S}(\varepsilon) = \{1+\exp[(\varepsilon-\mu_\text{S})/k_\text{B}T]\}^{-1}$ and
$f_\text{D}(\varepsilon) = f_\text{S}(\varepsilon+V_\text{SD})$ are the Fermi-Dirac distributions
at the source and drain, and $\mu_\text{S}$ is the source chemical potential. We
truncate the sum over $n$ to a few $k_n$ near $k_{n_0}$.

\section{Results}

In Fig.~\ref{fig2}(b) we consider
a narrow graphene well with $a=50$~nm and $b\rightarrow \infty$. The solution of
the graphene Dirac equation within the bandgap $\Delta_0$ region shows a
confined state. \cite{Guido2007} This state corresponds to the resonant spike
within the $\Delta_0$ region in Fig.~\ref{fig2}(b) for two barriers. For $N$
barriers the confined states hybridizes into $N-1$ states, leading to minibands
for large $N$; Figs.~\ref{fig2}(c) and \ref{fig2}(d). The minibands away from the
$\Delta_0$ region occur due to reflections at each interface. For finite bias
the minibands break into single resonant levels, Wannier-Stark ladders, as the
confined modes from each layer misalign; Fig.~\ref{fig2}(e). At the crossings of
Wannier-Stark ladders from distinct minibands the transmission increases due to
resonant tunneling.

\subsection*{NDR regimes} 
\begin{figure}
  \includegraphics[width=8cm]{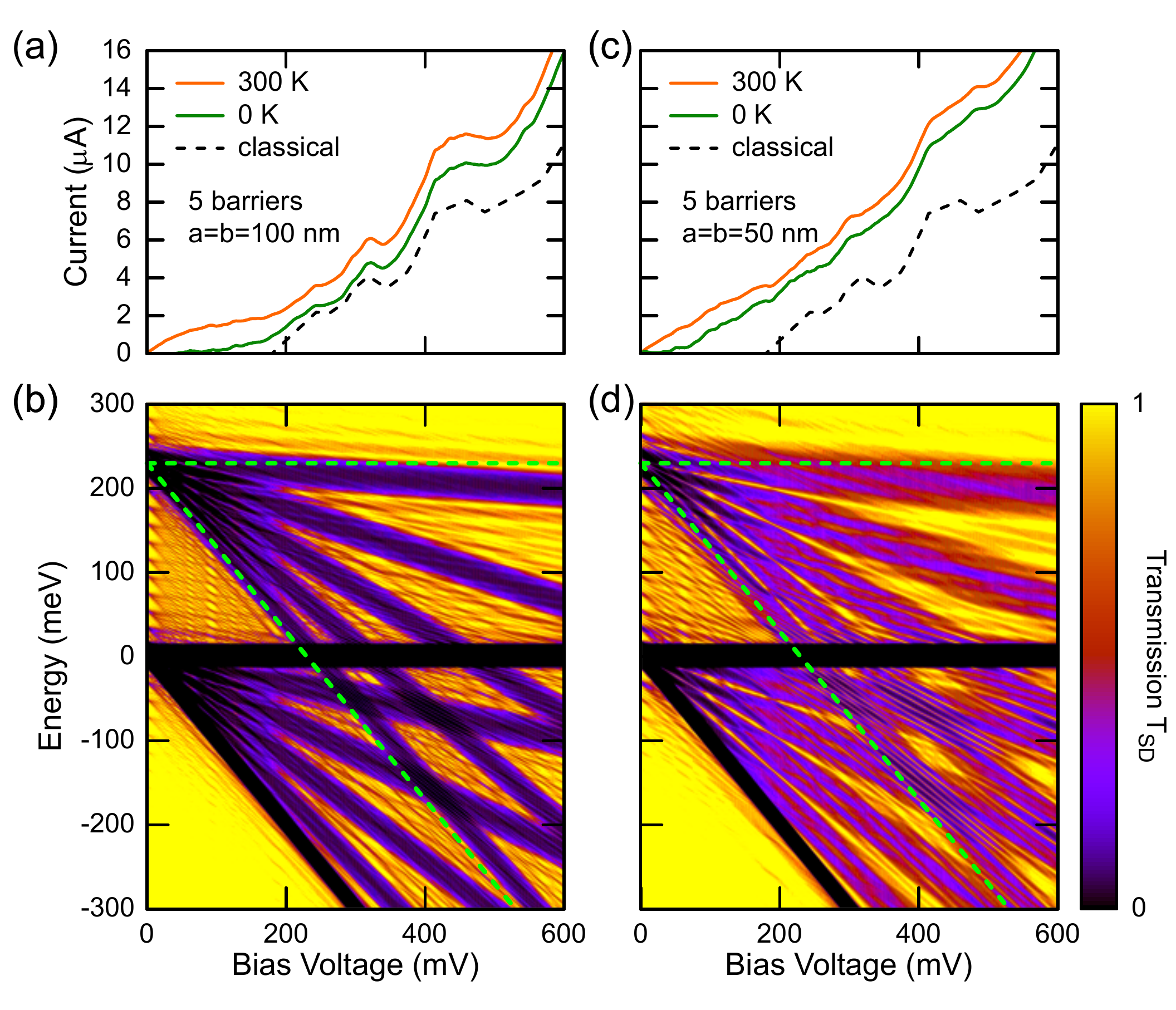}
  \caption{(Color online) Current and energy-voltage diagram of the transmission
coefficient for five-barrier superlattice with $a=b=100$~nm [(a) and (b)] and
$a=b=50$~nm [(c) and (d)]. The current-voltage characteristics are shown for
$T=300$~K and $0$~K.
  For wide barriers (a) and (b) the current follows closely the limiting
``classical'' case of $T_\text{SD}$ either 0 across bandgaps, or 1 otherwise (dashed
line).}
  \label{fig3}
\end{figure}

To contrast distinct NDR regimes in our system, we discuss
the current-voltage characteristics $I$-$V_\text{SD}$ and the energy-voltage
$T_\text{SD}$ diagram for the following three cases. We compare five-barrier
superlattices with (i) wide layers [Figs.~\ref{fig3}(a) and \ref{fig3}(b)] and (ii) narrow
layers [Figs.~\ref{fig3}(c) and \ref{fig3}(d)]. We then discuss (iii) a 20-barrier superlattice
with narrow layers; Fig.~\ref{fig4}. The dashed lines in the $T_\text{SD}$ diagrams
delimit the zero-temperature window of integration for $I_\text{SD}$, defined between
the source $\mu_\text{S} = 230$~meV and drain $\mu_\text{D} = \mu_\text{S} - V_\text{SD}$ chemical
potentials.

\subsubsection{``Classical'' regime} For wide layers, $a=b=100$~nm, tunneling
across bandgaps is strongly suppressed and the $T_\text{SD}$ diagram,
Fig.~\ref{fig3}(b), follows closely the diamond pattern in Fig.~\ref{fig1}(d).
For $eV_\text{SD} \lesssim \Delta\varepsilon_\text{F} = 230$~meV the current increases
monotonically as the barriers bandgaps misalign. At the coincidence region,
$eV_\text{SD} \gtrsim \Delta\varepsilon_\text{F} = 230$~meV, crossings of barrier and valley
bandgaps lead to the diamond pattern of finite and suppressed $T_\text{SD}$. This
alternation leads to the NDR near $V_\text{SD}=350$ and $450$~mV, in
Fig.~\ref{fig3}(a). The intensity of the NDR in this regime increases with the
layers width, as the tunneling across bandgaps becomes more suppressed. The
dashed curve in Fig.~\ref{fig3}(a) is calculated with the limiting case where
tunneling is completely suppressed across bandgaps, i.e., $T^{classical}_\text{SD} =
0$ across a bandgap, and $1$ otherwise. Note the similarity of the dashed
classical line with the exact $I_\text{SD}$ calculations in Fig.~\ref{fig3}(a).

For narrow layers, $a=b=50$~nm in Figs.~\ref{fig3}(c) and \ref{fig3}(d), the NDR due to
classical regime is absent as it requires strong tunneling suppression.
Interestingly, however, the $T_\text{SD}$ diagram of a few narrow layers clearly
shows the evolution of the zero-bias minibands into Wannier-Stark ladders with
increasing $V_\text{SD}$; Fig.~\ref{fig2}(e). The Wannier-Stark ladders remain as
individual transmission spikes while there is an overlap of barriers (or valley)
bandgaps. For $eV_\text{SD} > (N-1/2)\Delta_0$ this condition is violated, and the
tunneling across individual bandgaps dominate. At the crossings of barrier and
valley bandgaps, resonant effects are still visible in the $T_\text{SD}$ diagram as
stripes, corresponding to confined states between the overlapping band gaps; see
Fig.~\ref{fig3}(d) near $\varepsilon=-50$~meV and $V_\text{SD}=400$~mV.

\subsubsection{Miniband regime} 
\begin{figure}[ht]
  \includegraphics[width=8cm]{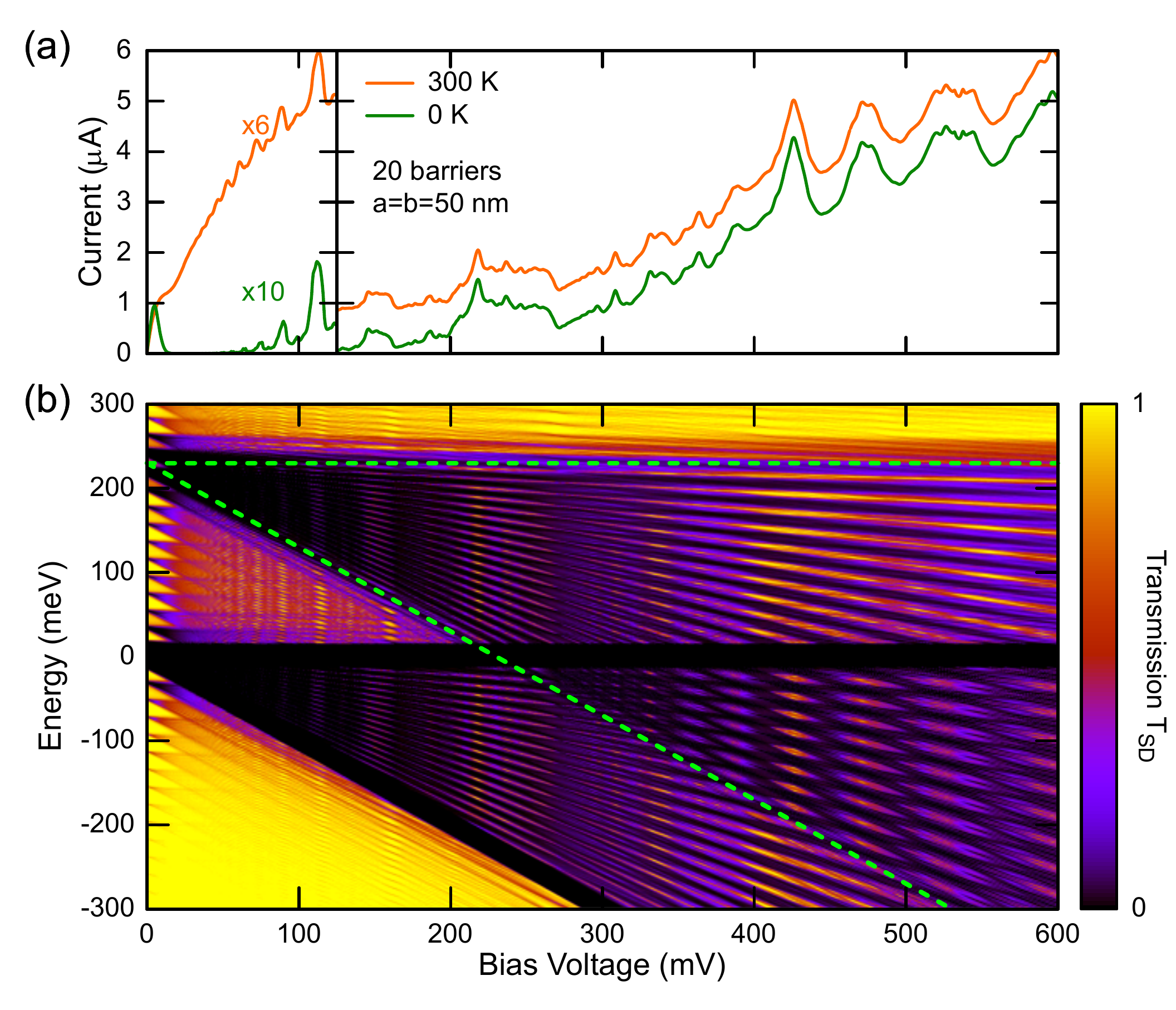}
  \caption{(Color online) (a) Current-voltage characteristics and (b) $T_\text{SD}$
diagram of a $20$-barrier aGNR superlattice with $a=b=50$~nm. In (a) the
currents for $0$ and $300$~K in the range $0 \leq V_\text{SD} \leq 125$~mV are
multiplied by $6$ and $10$, respectively, for clarity. As the voltage increases
the miniband near $230$~meV, Fig.~\ref{fig2}(e), breaks up as the resonant
levels misalign, leading to the pronounced spike near $10$~mV for $0$~K. Near
$50$~mV the resonant levels return as resonant crossings of Wannier-Stark ladder
rungs [see also Fig.~\ref{fig2}(a)]. At the crossings $T_\text{SD}$ increases,
showing current spikes at both $0$ and $300$~K for $V_\text{SD} < 230$~mV. For
$V_\text{SD} > 230$~mV the current spikes arise from crossings of rungs at the
coincidence region.}
  \label{fig4}
\end{figure}

Considering a larger number of barriers,
$N=20$ in Fig.~\ref{fig4}, the aligned resonant modes hybridize into
superlattice minibands; Fig.~\ref{fig2}. If $\mu_\text{S}$ is located within the
miniband, at low biases the current is dominated by the transmission across
these resonant modes. As the bias increases, the modes misalign breaking up the
miniband into Wannier-Stark ladders. For five barriers, Fig.~\ref{fig2}(a), the
rungs of the ladders shows nonresonant transmission peaks, and enhanced
resonant transmission at crossings of the rungs (see \textit{Wannier-Stark
ladder regime} below). For 20 barriers, transmission through nonresonant rungs
is strongly suppressed due to the larger number of bandgaps. At very low
voltages, Fig.~\ref{fig4}, the current initially increases with $V_\text{SD}$ as the
transport occurs through the miniband. Near $eV_\text{SD} \sim 10$~meV (of the order
of the miniband width) the miniband breaks up into the nonresonant rungs
suppressing the current, thus resulting in a pronounced current peak.

\subsubsection{Wannier-Stark ladder regime} With increasing bias, rungs from
Wannier-Stark ladders of distinct minibands cross, Fig.~\ref{fig2}(a), creating
new resonances through the superlattice layers. For 20 barriers, where
transmission from non-resonant rungs is strongly suppressed, the crossings show
sharp $T_\text{SD}$ stripes, e.g., at $V_\text{SD} = 75$, $110$, $150$, and $210$~mV;
Fig.~\ref{fig4}(b). Each of these stripes, and others with lower contrast at
smaller voltages, leads to current spikes in Fig.~\ref{fig4}(a). The spikes
broaden with increasing bias as the band gaps misalign. For $eV_\text{SD} >
\Delta\varepsilon_\text{F} = 230$~meV, the crossings of broadened Wannier-Stark ladders
from minibands near the barrier and valley bandgaps show diamond-shaped
structures in the $T_\text{SD}$ diagram, thus leading to a series of NDR spikes
similar to the classical regime.

\section{Conclusions}
We have found that three distinct regimes can lead to NDR in semiconducting aGNR
superlattices. (i) In the classical regime the NDR occurs as the bandgaps of
different layers cross with increasing $V_\text{SD}$. (ii) For narrow layers and very
low biases, $eV_\text{SD} \sim 10$~meV, the transport is dominated by the
resonant tunneling through the miniband, and the NDR occurs as the miniband
breaks into Wannier-Stark ladders with increasing bias. (iii) For higher bias
rungs of distinct ladders cross originating new resonances and current peaks.
Interestingly, due to the high mobility of the carriers, we obtain low bias NDR
peaks with high current densities.

\subsection*{Final remarks}
The predicted NDR effects reported here are strictly valid for ballistic electronic transport through ideal aGNR superlattices. For relatively clean systems, however, we expect detrimental effects such as those induced by disorder, impurities and structural defects\cite{abedpour2009conductance,cheraghchi2011metallic,han2010electron,saloriutta2011electron} to broaden the resonances in the $I$-$V$ curves, thus possibly reducing the peak-to-valley current ratios. Interestingly, a recent calculation for the electronic transport through a single-barrier defined on a zigzag-terminated graphene nanoribbon shows evidence for a transport gap despite the gapless spectrum of the edge states of the system. \cite{nakabayashi2009band} Therefore, we expect that a superlattice defined on a zigzag graphene nanoribbon should exhibit transport features similar to those of the armchair case investigated here. The effects of edge irregularities, strong disorder, and interactions (even at the Hartree level) lie beyond the scope of the present work and deserve further study.


\begin{acknowledgments}
We thank Bj\"orn Trauzettel, Saiful Khondaker, Volodymyr Turkowski, and Stephano
Chesi for useful discussions.
The authors acknowledge support from FAPESP, CNPq, Swiss NSF, and NCCR
Nanoscience. M.N.L. acknowledges support from NSF (Grant No. ECCS-0725514),
DARPA/MTO (Grant No. HR0011-08-1-0059), NSF (Grant No. ECCS-0901784), and AFOSR
(Grant No. FA9550-09-1-0450).
\end{acknowledgments}

\appendix 
\section{Transfer Matrix}
In this Appendix we detail the calculation of the transmission coefficient $T_\text{SD}$ through the nanoribbon superlattice via the transfer-matrix approach.
We describe the potential across the system as piecewise constant; Fig.~\ref{fig1}(b). In each layer the potential is a constant $V_j = V^\text{SL}_j - eV_\text{SD}x_j/L$. The superlattice potential $V_j^\text{SL}$ is 0 for valleys, and $V_\text{b} = 230$~mV for barriers (typical value obtained from Refs.~\onlinecite{GiovannettiPRL2008,VaninPRB2010,SalvadorPRL2010,Varykhalov2010}). The second term is the potential energy drop across the $j^\text{th}$ layers due to the electric field, where $x_j$ is the coordinate of the center of the layer $j$, and $L$ is the distance between the source and drain.

The solution of the Dirac equation in each layer $j$ ($j=\text{S}$ and $D$ for the source and drain, and an integer for the intermediate layers) is given by the plane-wave spinors\cite{KatsnelsonNature2006,CastroNeto2009} $\psi_{j,n}(x,y) = e^{ik_ny}\varphi_j(x)$. For convenience we write the $x$ component in a matrix form $\varphi_j(x) = M_j(x)\phi_j$, where the components of the spinor $\phi_j^\text{T} = (\alpha_j \;\; \beta_j)^T$ denote the coefficients of the outgoing and incoming plane waves. The matrix $M_j(x)$ is

\begin{equation}
 M_j(x) =
\left(
\begin{array}{c c}
e^{ik_x^{(j)}x} & e^{-ik_x^{(j)}x}\\
s_je^{ik_x^{(j)}x+i\theta_{j,n}} & -s_je^{-ik_x^{(j)}x-i\theta_{j,n}}
\end{array}
\right).
\end{equation}
The eigenenergies in each layer are $\varepsilon_{j,n} = V_j + s_j \hbar v_\text{f}\sqrt{(k_x^{(j)})^2 + k_n^2}$, with $s_j = +1$ for the conduction band and $s_j = -1$ for the valence band, $k_x^{(j)}$ is the longitudinal momentum in layer $j$, $k_n$ is the quantized transversal momentum (conserved through the system), and $\theta_{j,n} = \tan^{-1}(k_n/k_x^{(j)})$.

The continuity of the spinors at the interfaces yields 
$\varphi_j(x_{j,j+1}) = \varphi_{j+1}(x_{j,j+1})$, where $x_{j,j+1}$ is the position of the interface between the layers $j$ and $j+1$. Applying this matching throughout the system, we obtain a $2\times 2$ matrix equation connecting the coefficients from source and drain $\phi_{\text{S}} = T_{\text{M}} \phi_{\text{D}}$, where $T_\text{M}$ is the transfer matrix given by

\begin{equation}
 T_{\text{M}} = \prod_j M_j^{-1}(x_{j,j+1})M_{j+1}(x_{j,j+1}).
\end{equation}

The definition of the reflected and transmitted waves depends on the sign of the electron energy at source $s_{\text{S}}$ and drain $s_\text{D}$, such that the source and drain coefficients are given by

\begin{eqnarray}
 \phi_\text{S}^\text{T} &=& \left\{
 \begin{array}{c}
  (1 \;\; r), \text{ if } s_\text{S} = +1,\\
  (r \;\; 1), \text{ if } s_\text{S} = -1,
 \end{array}
\right.\\
 \phi_\text{D}^\text{T} &=& \left\{
 \begin{array}{c}
  (t \;\; 0), \text{ if } s_\text{D} = +1,\\
  (0 \;\; t), \text{ if } s_\text{D} = -1.
 \end{array}
\right.
\end{eqnarray}

From the graphene Dirac Hamiltonian, the current density reads $J^{(j)}_x = 4ev_\text{f} \varphi_j^\dagger(x)\sigma_x\varphi_j(x)$. At the stationary regime the current flow at source and drain is the same, requiring the match $J^\text{S}_x = J^\text{D}_x$, from which we identify the transmission coefficient $T_\text{SD}$,

\begin{equation}
 T_\text{SD}(\varepsilon, k_n, V_\text{SD}) = |t|^2\dfrac{\cos\theta_\text{D}}{\cos\theta_\text{S}}.
\end{equation}
This transmission coefficient as a function of the energy reveals the electronic structure of the system, in which the confined modes in between the layers show up as resonant spikes and minibands; Fig.~\ref{fig2}.


\begin{thebibliography}{42}
\expandafter\ifx\csname natexlab\endcsname\relax\def\natexlab#1{#1}\fi
\expandafter\ifx\csname bibnamefont\endcsname\relax
  \def\bibnamefont#1{#1}\fi
\expandafter\ifx\csname bibfnamefont\endcsname\relax
  \def\bibfnamefont#1{#1}\fi
\expandafter\ifx\csname citenamefont\endcsname\relax
  \def\citenamefont#1{#1}\fi
\expandafter\ifx\csname url\endcsname\relax
  \def\url#1{\texttt{#1}}\fi
\expandafter\ifx\csname urlprefix\endcsname\relax\def\urlprefix{URL }\fi
\providecommand{\bibinfo}[2]{#2}
\providecommand{\eprint}[2][]{\url{#2}}

\bibitem[{\citenamefont{Novoselov et~al.}(2005)\citenamefont{Novoselov, Geim,
  Morozov, Jiang, Katsnelson, Grigorieva, Dubonos, and
  Firsov}}]{novoselov2005two}
\bibinfo{author}{\bibfnamefont{K.~S.} \bibnamefont{Novoselov}},
  \bibinfo{author}{\bibfnamefont{A.~K.} \bibnamefont{Geim}},
  \bibinfo{author}{\bibfnamefont{S.~V.} \bibnamefont{Morozov}},
  \bibinfo{author}{\bibfnamefont{D.}~\bibnamefont{Jiang}},
  \bibinfo{author}{\bibfnamefont{M.~I.} \bibnamefont{Katsnelson}},
  \bibinfo{author}{\bibfnamefont{I.~V.} \bibnamefont{Grigorieva}},
  \bibinfo{author}{\bibfnamefont{S.~V.} \bibnamefont{Dubonos}},
  \bibnamefont{and} \bibinfo{author}{\bibfnamefont{A.~A.}
  \bibnamefont{Firsov}}, \bibinfo{journal}{Nature (London)}
  \textbf{\bibinfo{volume}{438}}, \bibinfo{pages}{197} (\bibinfo{year}{2005}).

\bibitem[{\citenamefont{Katsnelson et~al.}(2006)\citenamefont{Katsnelson,
  Novoselov, and Gaim}}]{KatsnelsonNature2006}
\bibinfo{author}{\bibfnamefont{M.~I.} \bibnamefont{Katsnelson}},
  \bibinfo{author}{\bibfnamefont{K.~S.} \bibnamefont{Novoselov}},
  \bibnamefont{and} \bibinfo{author}{\bibfnamefont{A.~K.} \bibnamefont{Gaim}},
  \bibinfo{journal}{Nat. Phys.} \textbf{\bibinfo{volume}{2}},
  \bibinfo{pages}{620} (\bibinfo{year}{2006}).

\bibitem[{\citenamefont{Neto et~al.}(2009)\citenamefont{Neto, Guinea, Peres,
  Novoselov, and Gaim}}]{CastroNeto2009}
\bibinfo{author}{\bibfnamefont{A.~H.~C.} \bibnamefont{Neto}},
  \bibinfo{author}{\bibfnamefont{F.}~\bibnamefont{Guinea}},
  \bibinfo{author}{\bibfnamefont{N.~M.~R.} \bibnamefont{Peres}},
  \bibinfo{author}{\bibfnamefont{K.~S.} \bibnamefont{Novoselov}},
  \bibnamefont{and} \bibinfo{author}{\bibfnamefont{A.~K.} \bibnamefont{Gaim}},
  \bibinfo{journal}{Rev. Mod. Phys.} \textbf{\bibinfo{volume}{81}},
  \bibinfo{pages}{109} (\bibinfo{year}{2009}).

\bibitem[{\citenamefont{Avouris}(2010)}]{avouris2010graphene}
\bibinfo{author}{\bibfnamefont{P.}~\bibnamefont{Avouris}},
  \bibinfo{journal}{Nano Lett.} \textbf{\bibinfo{volume}{10}},
  \bibinfo{pages}{4285} (\bibinfo{year}{2010}).

\bibitem[{\citenamefont{Yu and Cardona}(2005)}]{Cardona_book}
\bibinfo{author}{\bibfnamefont{P.~Y.} \bibnamefont{Yu}} \bibnamefont{and}
  \bibinfo{author}{\bibfnamefont{M.}~\bibnamefont{Cardona}},
  \emph{\bibinfo{title}{Fundamentals of Semiconductors}}
  (\bibinfo{publisher}{Springer, Berlin}, \bibinfo{year}{2005}).

\bibitem[{\citenamefont{Mortazawi et~al.}(1989)\citenamefont{Mortazawi, Kesan,
  Neikirk, and Itoh}}]{Mortazawi}
\bibinfo{author}{\bibfnamefont{A.}~\bibnamefont{Mortazawi}},
  \bibinfo{author}{\bibfnamefont{V.}~\bibnamefont{Kesan}},
  \bibinfo{author}{\bibfnamefont{D.}~\bibnamefont{Neikirk}}, \bibnamefont{and}
  \bibinfo{author}{\bibfnamefont{T.}~\bibnamefont{Itoh}}, in
  \emph{\bibinfo{booktitle}{Microwave Conference, 1989. 19th European}}
  (\bibinfo{year}{1989}), pp. \bibinfo{pages}{715--718}.

\bibitem[{\citenamefont{Sollner et~al.}(1988)\citenamefont{Sollner, Brown,
  Goodhue, and Correa}}]{Sollner}
\bibinfo{author}{\bibfnamefont{T.~C. L.~G.} \bibnamefont{Sollner}},
  \bibinfo{author}{\bibfnamefont{E.~R.} \bibnamefont{Brown}},
  \bibinfo{author}{\bibfnamefont{W.~D.} \bibnamefont{Goodhue}},
  \bibnamefont{and} \bibinfo{author}{\bibfnamefont{C.~A.}
  \bibnamefont{Correa}}, \bibinfo{journal}{J. Appl. Phys.}
  \textbf{\bibinfo{volume}{64}}, \bibinfo{pages}{4248} (\bibinfo{year}{1988}).

\bibitem[{\citenamefont{Sze and Ng}(2007)}]{sze2007physics}
\bibinfo{author}{\bibfnamefont{S.~M.} \bibnamefont{Sze}} \bibnamefont{and}
  \bibinfo{author}{\bibfnamefont{K.~K.} \bibnamefont{Ng}},
  \emph{\bibinfo{title}{Physics of Semiconductor Devices}}
  (\bibinfo{publisher}{Wiley-Interscience}, \bibinfo{address}{New York},
  \bibinfo{year}{2007}).

\bibitem[{\citenamefont{Tsu}(1973)}]{Tsu1973}
\bibinfo{author}{\bibfnamefont{R.}~\bibnamefont{Tsu}}, \bibinfo{journal}{Appl.
  Phys. Lett.} \textbf{\bibinfo{volume}{22}}, \bibinfo{pages}{562}
  (\bibinfo{year}{1973}).

\bibitem[{\citenamefont{Sollner}(1983)}]{SollnerAPL1983}
\bibinfo{author}{\bibfnamefont{T.~C. L.~G.} \bibnamefont{Sollner}},
  \bibinfo{journal}{Appl. Phys. Lett.} \textbf{\bibinfo{volume}{43}},
  \bibinfo{pages}{588} (\bibinfo{year}{1983}).

\bibitem[{\citenamefont{Iogansen}(1964)}]{Iogansen1964}
\bibinfo{author}{\bibfnamefont{L.~V.} \bibnamefont{Iogansen}},
  \bibinfo{journal}{Sov. Phys. JETP} \textbf{\bibinfo{volume}{18}},
  \bibinfo{pages}{146} (\bibinfo{year}{1964}).

\bibitem[{\citenamefont{Esaki and Tsu}(1970)}]{Esaki_Tsu}
\bibinfo{author}{\bibfnamefont{L.}~\bibnamefont{Esaki}} \bibnamefont{and}
  \bibinfo{author}{\bibfnamefont{R.}~\bibnamefont{Tsu}}, \bibinfo{journal}{IBM
  J. Res. Develop.} \textbf{\bibinfo{volume}{14}}, \bibinfo{pages}{61}
  (\bibinfo{year}{1970}).

\bibitem[{\citenamefont{Tsu}(2005)}]{Tsu_book}
\bibinfo{author}{\bibfnamefont{R.}~\bibnamefont{Tsu}},
  \emph{\bibinfo{title}{Superlattice to Nanoelectronics}}
  (\bibinfo{publisher}{Elsevier}, \bibinfo{address}{Amsterdam},
  \bibinfo{year}{2005}).

\bibitem[{\citenamefont{Goldman et~al.}(1987)\citenamefont{Goldman, Tsui, and
  Cunningham}}]{Goldman1987}
\bibinfo{author}{\bibfnamefont{V.~J.} \bibnamefont{Goldman}},
  \bibinfo{author}{\bibfnamefont{D.~C.} \bibnamefont{Tsui}}, \bibnamefont{and}
  \bibinfo{author}{\bibfnamefont{J.~E.} \bibnamefont{Cunningham}},
  \bibinfo{journal}{Phys. Rev. Lett.} \textbf{\bibinfo{volume}{58}},
  \bibinfo{pages}{1256} (\bibinfo{year}{1987}).

\bibitem[{\citenamefont{Bai and Zhang}(2007)}]{bai2007klein}
\bibinfo{author}{\bibfnamefont{C.}~\bibnamefont{Bai}} \bibnamefont{and}
  \bibinfo{author}{\bibfnamefont{X.}~\bibnamefont{Zhang}},
  \bibinfo{journal}{Physical Review B} \textbf{\bibinfo{volume}{76}},
  \bibinfo{pages}{075430} (\bibinfo{year}{2007}).

\bibitem[{\citenamefont{Brey and Fertig}(2009)}]{brey2009emerging}
\bibinfo{author}{\bibfnamefont{L.}~\bibnamefont{Brey}} \bibnamefont{and}
  \bibinfo{author}{\bibfnamefont{H.~A.} \bibnamefont{Fertig}},
  \bibinfo{journal}{Physical Review Letters} \textbf{\bibinfo{volume}{103}},
  \bibinfo{pages}{46809} (\bibinfo{year}{2009}).

\bibitem[{\citenamefont{Park et~al.}(2009)\citenamefont{Park, Son, Yang, Cohen,
  and Louie}}]{park2009landau}
\bibinfo{author}{\bibfnamefont{C.~H.} \bibnamefont{Park}},
  \bibinfo{author}{\bibfnamefont{Y.~W.} \bibnamefont{Son}},
  \bibinfo{author}{\bibfnamefont{L.}~\bibnamefont{Yang}},
  \bibinfo{author}{\bibfnamefont{M.~L.} \bibnamefont{Cohen}}, \bibnamefont{and}
  \bibinfo{author}{\bibfnamefont{S.~G.} \bibnamefont{Louie}},
  \bibinfo{journal}{Physical Review Letters} \textbf{\bibinfo{volume}{103}},
  \bibinfo{pages}{46808} (\bibinfo{year}{2009}).

\bibitem[{\citenamefont{Barbier et~al.}(2010)\citenamefont{Barbier,
  Vasilopoulos, and Peeters}}]{barbier2010extra}
\bibinfo{author}{\bibfnamefont{M.}~\bibnamefont{Barbier}},
  \bibinfo{author}{\bibfnamefont{P.}~\bibnamefont{Vasilopoulos}},
  \bibnamefont{and} \bibinfo{author}{\bibfnamefont{F.~M.}
  \bibnamefont{Peeters}}, \bibinfo{journal}{Physical Review B}
  \textbf{\bibinfo{volume}{81}}, \bibinfo{pages}{075438}
  (\bibinfo{year}{2010}).

\bibitem[{\citenamefont{Stojanovi{\'c}
  et~al.}(2010)\citenamefont{Stojanovi{\'c}, Vukmirovi{\'c}, and
  Bruder}}]{stojanovic2010polaronic}
\bibinfo{author}{\bibfnamefont{V.~M.} \bibnamefont{Stojanovi{\'c}}},
  \bibinfo{author}{\bibfnamefont{N.}~\bibnamefont{Vukmirovi{\'c}}},
  \bibnamefont{and} \bibinfo{author}{\bibfnamefont{C.}~\bibnamefont{Bruder}},
  \bibinfo{journal}{Physical Review B} \textbf{\bibinfo{volume}{82}},
  \bibinfo{pages}{165410} (\bibinfo{year}{2010}).

\bibitem[{\citenamefont{Burset et~al.}(2011)\citenamefont{Burset, Yeyati, Brey,
  and Fertig}}]{burset2011transport}
\bibinfo{author}{\bibfnamefont{P.}~\bibnamefont{Burset}},
  \bibinfo{author}{\bibfnamefont{A.~L.} \bibnamefont{Yeyati}},
  \bibinfo{author}{\bibfnamefont{L.}~\bibnamefont{Brey}}, \bibnamefont{and}
  \bibinfo{author}{\bibfnamefont{H.~A.} \bibnamefont{Fertig}},
  \bibinfo{journal}{Physical Review B} \textbf{\bibinfo{volume}{83}},
  \bibinfo{pages}{195434} (\bibinfo{year}{2011}).

\bibitem[{\citenamefont{Guo et~al.}(2011)\citenamefont{Guo, Liu, and
  Li}}]{guo2011conductance}
\bibinfo{author}{\bibfnamefont{X.}~\bibnamefont{Guo}},
  \bibinfo{author}{\bibfnamefont{D.}~\bibnamefont{Liu}}, \bibnamefont{and}
  \bibinfo{author}{\bibfnamefont{Y.}~\bibnamefont{Li}},
  \bibinfo{journal}{Applied Physics Letters} \textbf{\bibinfo{volume}{98}},
  \bibinfo{pages}{242101} (\bibinfo{year}{2011}).

\bibitem[{\citenamefont{Jiang et~al.}(2011)\citenamefont{Jiang, Wang, and
  Wang}}]{jiang2011minimum}
\bibinfo{author}{\bibfnamefont{J.}~\bibnamefont{Jiang}},
  \bibinfo{author}{\bibfnamefont{J.}~\bibnamefont{Wang}}, \bibnamefont{and}
  \bibinfo{author}{\bibfnamefont{B.}~\bibnamefont{Wang}},
  \bibinfo{journal}{Applied Physics Letters} \textbf{\bibinfo{volume}{99}},
  \bibinfo{pages}{043109} (\bibinfo{year}{2011}).

\bibitem[{\citenamefont{Abedpour et~al.}(2009)\citenamefont{Abedpour,
  Esmailpour, Asgari, and Tabar}}]{abedpour2009conductance}
\bibinfo{author}{\bibfnamefont{N.}~\bibnamefont{Abedpour}},
  \bibinfo{author}{\bibfnamefont{A.}~\bibnamefont{Esmailpour}},
  \bibinfo{author}{\bibfnamefont{R.}~\bibnamefont{Asgari}}, \bibnamefont{and}
  \bibinfo{author}{\bibfnamefont{M.~R.} \bibnamefont{Tabar}},
  \bibinfo{journal}{Physical Review B} \textbf{\bibinfo{volume}{79}},
  \bibinfo{pages}{165412} (\bibinfo{year}{2009}).

\bibitem[{\citenamefont{Cheraghchi et~al.}(2011)\citenamefont{Cheraghchi,
  Irani, Fazeli, and Asgari}}]{cheraghchi2011metallic}
\bibinfo{author}{\bibfnamefont{H.}~\bibnamefont{Cheraghchi}},
  \bibinfo{author}{\bibfnamefont{A.~H.} \bibnamefont{Irani}},
  \bibinfo{author}{\bibfnamefont{S.~M.} \bibnamefont{Fazeli}},
  \bibnamefont{and} \bibinfo{author}{\bibfnamefont{R.}~\bibnamefont{Asgari}},
  \bibinfo{journal}{Physical Review B} \textbf{\bibinfo{volume}{83}},
  \bibinfo{pages}{235430} (\bibinfo{year}{2011}).

\bibitem[{\citenamefont{Giovannetti et~al.}(2008)\citenamefont{Giovannetti,
  Khomyakov, Brocks, Karpan, {van den Brink}, and Kelly}}]{GiovannettiPRL2008}
\bibinfo{author}{\bibfnamefont{G.}~\bibnamefont{Giovannetti}},
  \bibinfo{author}{\bibfnamefont{P.~A.} \bibnamefont{Khomyakov}},
  \bibinfo{author}{\bibfnamefont{G.}~\bibnamefont{Brocks}},
  \bibinfo{author}{\bibfnamefont{V.~M.} \bibnamefont{Karpan}},
  \bibinfo{author}{\bibfnamefont{J.}~\bibnamefont{{van den Brink}}},
  \bibnamefont{and} \bibinfo{author}{\bibfnamefont{P.~J.} \bibnamefont{Kelly}},
  \bibinfo{journal}{Phys. Rev. Lett.} \textbf{\bibinfo{volume}{101}},
  \bibinfo{pages}{026803} (\bibinfo{year}{2008}).

\bibitem[{\citenamefont{Vanin et~al.}(2010)\citenamefont{Vanin, Mortensen,
  Kelkkanen, Garcia-Lastra, Thygesen, and Jacobsen}}]{VaninPRB2010}
\bibinfo{author}{\bibfnamefont{M.}~\bibnamefont{Vanin}},
  \bibinfo{author}{\bibfnamefont{J.~J.} \bibnamefont{Mortensen}},
  \bibinfo{author}{\bibfnamefont{A.~K.} \bibnamefont{Kelkkanen}},
  \bibinfo{author}{\bibfnamefont{J.~M.} \bibnamefont{Garcia-Lastra}},
  \bibinfo{author}{\bibfnamefont{K.~S.} \bibnamefont{Thygesen}},
  \bibnamefont{and} \bibinfo{author}{\bibfnamefont{K.~W.}
  \bibnamefont{Jacobsen}}, \bibinfo{journal}{Phys. Rev. B}
  \textbf{\bibinfo{volume}{81}}, \bibinfo{pages}{081408(R)}
  (\bibinfo{year}{2010}).

\bibitem[{\citenamefont{Barraza-Lopez et~al.}(2010)\citenamefont{Barraza-Lopez,
  Vanevi{\'c}, Kindermann, and Chou}}]{SalvadorPRL2010}
\bibinfo{author}{\bibfnamefont{S.}~\bibnamefont{Barraza-Lopez}},
  \bibinfo{author}{\bibfnamefont{M.}~\bibnamefont{Vanevi{\'c}}},
  \bibinfo{author}{\bibfnamefont{M.}~\bibnamefont{Kindermann}},
  \bibnamefont{and} \bibinfo{author}{\bibfnamefont{M.~Y.} \bibnamefont{Chou}},
  \bibinfo{journal}{Phys. Rev. Lett.} \textbf{\bibinfo{volume}{104}},
  \bibinfo{pages}{076807} (\bibinfo{year}{2010}).

\bibitem[{\citenamefont{Varykhalov et~al.}(2010)\citenamefont{Varykhalov,
  Scholz, Kim, and Rader}}]{Varykhalov2010}
\bibinfo{author}{\bibfnamefont{A.}~\bibnamefont{Varykhalov}},
  \bibinfo{author}{\bibfnamefont{M.~R.} \bibnamefont{Scholz}},
  \bibinfo{author}{\bibfnamefont{T.~K.} \bibnamefont{Kim}}, \bibnamefont{and}
  \bibinfo{author}{\bibfnamefont{O.}~\bibnamefont{Rader}},
  \bibinfo{journal}{Phys. Rev. B} \textbf{\bibinfo{volume}{82}},
  \bibinfo{pages}{121101} (\bibinfo{year}{2010}).

\bibitem[{\citenamefont{Wang et~al.}(2008)\citenamefont{Wang, Li, Shi, Wang,
  Yang, Hou, and Chen}}]{WangAPL2008}
\bibinfo{author}{\bibfnamefont{Z.~F.} \bibnamefont{Wang}},
  \bibinfo{author}{\bibfnamefont{Q.}~\bibnamefont{Li}},
  \bibinfo{author}{\bibfnamefont{Q.~W.} \bibnamefont{Shi}},
  \bibinfo{author}{\bibfnamefont{X.}~\bibnamefont{Wang}},
  \bibinfo{author}{\bibfnamefont{J.}~\bibnamefont{Yang}},
  \bibinfo{author}{\bibfnamefont{J.~G.} \bibnamefont{Hou}}, \bibnamefont{and}
  \bibinfo{author}{\bibfnamefont{J.}~\bibnamefont{Chen}},
  \bibinfo{journal}{Appl. Phys. Lett.} \textbf{\bibinfo{volume}{92}},
  \bibinfo{pages}{133114} (\bibinfo{year}{2008}).

\bibitem[{\citenamefont{Do and Dollfus}(2010)}]{NamDoJAP2010}
\bibinfo{author}{\bibfnamefont{V.~N.} \bibnamefont{Do}} \bibnamefont{and}
  \bibinfo{author}{\bibfnamefont{P.}~\bibnamefont{Dollfus}},
  \bibinfo{journal}{J. Appl. Phys.} \textbf{\bibinfo{volume}{107}},
  \bibinfo{pages}{063705} (\bibinfo{year}{2010}).

\bibitem{footnoteDragoman}
A recent work (Ref.~\onlinecite{DragomanAPL2007}) has claimed that low-bias NDR can be achieved with a single barrier in a infinite graphene sheet. This, however, has been disputed in Refs.~\onlinecite{NamDoCommentAPL2007,NamDoJAP2008}.

\bibitem[{\citenamefont{Ren et~al.}(2009)\citenamefont{Ren, li, Luo, and
  Yang}}]{RenAPL2009}
\bibinfo{author}{\bibfnamefont{H.}~\bibnamefont{Ren}},
  \bibinfo{author}{\bibfnamefont{Q.-X.} \bibnamefont{li}},
  \bibinfo{author}{\bibfnamefont{Y.}~\bibnamefont{Luo}}, \bibnamefont{and}
  \bibinfo{author}{\bibfnamefont{J.}~\bibnamefont{Yang}},
  \bibinfo{journal}{Appl. Phys. Lett.} \textbf{\bibinfo{volume}{94}},
  \bibinfo{pages}{173110} (\bibinfo{year}{2009}).

\bibitem[{\citenamefont{Habib et~al.}(2011)\citenamefont{Habib, Zahid, and
  Lake}}]{habib2011negative}
\bibinfo{author}{\bibfnamefont{K.}~\bibnamefont{Habib}},
  \bibinfo{author}{\bibfnamefont{F.}~\bibnamefont{Zahid}}, \bibnamefont{and}
  \bibinfo{author}{\bibfnamefont{R.}~\bibnamefont{Lake}},
  \bibinfo{journal}{Applied Physics Letters} \textbf{\bibinfo{volume}{98}},
  \bibinfo{pages}{192112} (\bibinfo{year}{2011}).

\bibitem[{\citenamefont{Fang et~al.}(2011)\citenamefont{Fang, Wang, Chen, Yan,
  Song, and Wang}}]{fang2011strain}
\bibinfo{author}{\bibfnamefont{H.}~\bibnamefont{Fang}},
  \bibinfo{author}{\bibfnamefont{R.}~\bibnamefont{Wang}},
  \bibinfo{author}{\bibfnamefont{S.}~\bibnamefont{Chen}},
  \bibinfo{author}{\bibfnamefont{M.}~\bibnamefont{Yan}},
  \bibinfo{author}{\bibfnamefont{X.}~\bibnamefont{Song}}, \bibnamefont{and}
  \bibinfo{author}{\bibfnamefont{B.}~\bibnamefont{Wang}},
  \bibinfo{journal}{Applied Physics Letters} \textbf{\bibinfo{volume}{98}},
  \bibinfo{pages}{082108} (\bibinfo{year}{2011}).

\bibitem[{\citenamefont{Krahne et~al.}(2002)\citenamefont{Krahne, Yacoby,
  Shtrikman, Bar-Joseph, Dadosh, and Sperling}}]{Krahne2002}
\bibinfo{author}{\bibfnamefont{R.}~\bibnamefont{Krahne}},
  \bibinfo{author}{\bibfnamefont{A.}~\bibnamefont{Yacoby}},
  \bibinfo{author}{\bibfnamefont{H.}~\bibnamefont{Shtrikman}},
  \bibinfo{author}{\bibfnamefont{I.}~\bibnamefont{Bar-Joseph}},
  \bibinfo{author}{\bibfnamefont{T.}~\bibnamefont{Dadosh}}, \bibnamefont{and}
  \bibinfo{author}{\bibfnamefont{J.}~\bibnamefont{Sperling}},
  \bibinfo{journal}{Appl. Phys. Lett.} \textbf{\bibinfo{volume}{81}},
  \bibinfo{pages}{730} (\bibinfo{year}{2002}).

\bibitem[{\citenamefont{Wallace}(1947)}]{Wallace1947}
\bibinfo{author}{\bibfnamefont{P.~R.} \bibnamefont{Wallace}},
  \bibinfo{journal}{Phys. Rev.} \textbf{\bibinfo{volume}{71}},
  \bibinfo{pages}{622} (\bibinfo{year}{1947}).

\bibitem[{\citenamefont{Brey and Fertig}(2006)}]{BreyFertigPRB2006}
\bibinfo{author}{\bibfnamefont{L.}~\bibnamefont{Brey}} \bibnamefont{and}
  \bibinfo{author}{\bibfnamefont{H.~A.} \bibnamefont{Fertig}},
  \bibinfo{journal}{Phys. Rev. B} \textbf{\bibinfo{volume}{73}},
  \bibinfo{pages}{235411} (\bibinfo{year}{2006}).

\bibitem[{\citenamefont{Datta}(1997)}]{Datta95}
\bibinfo{author}{\bibfnamefont{S.}~\bibnamefont{Datta}},
  \emph{\bibinfo{title}{Electronic Transport in Mesoscopic Systems}}
  (\bibinfo{publisher}{Cambridge University Press},
  \bibinfo{address}{Cambridge, England}, \bibinfo{year}{1997}).

\bibitem[{\citenamefont{Blanter and Buttiker}(2000)}]{BlanterButtiker2000}
\bibinfo{author}{\bibfnamefont{Y.~M.} \bibnamefont{Blanter}} \bibnamefont{and}
  \bibinfo{author}{\bibfnamefont{M.}~\bibnamefont{Buttiker}},
  \bibinfo{journal}{Phys. Rep.} \textbf{\bibinfo{volume}{336}},
  \bibinfo{pages}{1} (\bibinfo{year}{2000}).

\bibitem[{\citenamefont{Trauzettel et~al.}(2007)\citenamefont{Trauzettel,
  Bulaev, Loss, and Burkard}}]{Guido2007}
\bibinfo{author}{\bibfnamefont{B.}~\bibnamefont{Trauzettel}},
  \bibinfo{author}{\bibfnamefont{D.~V.} \bibnamefont{Bulaev}},
  \bibinfo{author}{\bibfnamefont{D.}~\bibnamefont{Loss}}, \bibnamefont{and}
  \bibinfo{author}{\bibfnamefont{G.}~\bibnamefont{Burkard}},
  \bibinfo{journal}{Nat. Phys.} \textbf{\bibinfo{volume}{3}},
  \bibinfo{pages}{192} (\bibinfo{year}{2007}).

\bibitem[{\citenamefont{Han et~al.}(2010)\citenamefont{Han, Brant, and
  Kim}}]{han2010electron}
\bibinfo{author}{\bibfnamefont{M.}~\bibnamefont{Han}},
  \bibinfo{author}{\bibfnamefont{J.}~\bibnamefont{Brant}}, \bibnamefont{and}
  \bibinfo{author}{\bibfnamefont{P.}~\bibnamefont{Kim}},
  \bibinfo{journal}{Physical review letters} \textbf{\bibinfo{volume}{104}},
  \bibinfo{pages}{56801} (\bibinfo{year}{2010}).

\bibitem[{\citenamefont{Saloriutta et~al.}(2011)\citenamefont{Saloriutta,
  Hancock, K{\"a}rkk{\"a}inen, K{\"a}rkk{\"a}inen, Puska, and
  Jauho}}]{saloriutta2011electron}
\bibinfo{author}{\bibfnamefont{K.}~\bibnamefont{Saloriutta}},
  \bibinfo{author}{\bibfnamefont{Y.}~\bibnamefont{Hancock}},
  \bibinfo{author}{\bibfnamefont{A.}~\bibnamefont{K{\"a}rkk{\"a}inen}},
  \bibinfo{author}{\bibfnamefont{L.}~\bibnamefont{K{\"a}rkk{\"a}inen}},
  \bibinfo{author}{\bibfnamefont{M.~J.} \bibnamefont{Puska}}, \bibnamefont{and}
  \bibinfo{author}{\bibfnamefont{A.~P.} \bibnamefont{Jauho}},
  \bibinfo{journal}{Physical Review B} \textbf{\bibinfo{volume}{83}},
  \bibinfo{pages}{205125} (\bibinfo{year}{2011}).

\bibitem[{\citenamefont{Nakabayashi et~al.}(2009)\citenamefont{Nakabayashi,
  Yamamoto, and Kurihara}}]{nakabayashi2009band}
\bibinfo{author}{\bibfnamefont{J.}~\bibnamefont{Nakabayashi}},
  \bibinfo{author}{\bibfnamefont{D.}~\bibnamefont{Yamamoto}}, \bibnamefont{and}
  \bibinfo{author}{\bibfnamefont{S.}~\bibnamefont{Kurihara}},
  \bibinfo{journal}{Physical Review Letters} \textbf{\bibinfo{volume}{102}},
  \bibinfo{pages}{66803} (\bibinfo{year}{2009}).

\bibitem[{\citenamefont{Dragoman and Dragoman}(2007)}]{DragomanAPL2007}
\bibinfo{author}{\bibfnamefont{D.}~\bibnamefont{Dragoman}} \bibnamefont{and}
  \bibinfo{author}{\bibfnamefont{M.}~\bibnamefont{Dragoman}},
  \bibinfo{journal}{Appl. Phys. Lett.} \textbf{\bibinfo{volume}{90}},
  \bibinfo{pages}{143111} (\bibinfo{year}{2007}).

\bibitem[{\citenamefont{Do}(2008)}]{NamDoCommentAPL2007}
\bibinfo{author}{\bibfnamefont{V.~N.} \bibnamefont{Do}},
  \bibinfo{journal}{Appl. Phys. Lett.} \textbf{\bibinfo{volume}{92}},
  \bibinfo{pages}{216101} (\bibinfo{year}{2008}).

\bibitem[{\citenamefont{Do et~al.}(2008)\citenamefont{Do, Nguyen, Dollfus, and
  Bournel}}]{NamDoJAP2008}
\bibinfo{author}{\bibfnamefont{V.~N.} \bibnamefont{Do}},
  \bibinfo{author}{\bibfnamefont{V.~H.} \bibnamefont{Nguyen}},
  \bibinfo{author}{\bibfnamefont{P.}~\bibnamefont{Dollfus}}, \bibnamefont{and}
  \bibinfo{author}{\bibfnamefont{A.}~\bibnamefont{Bournel}},
  \bibinfo{journal}{J. Appl. Phys.} \textbf{\bibinfo{volume}{104}},
  \bibinfo{pages}{063708} (\bibinfo{year}{2008}).
  
\end{thebibliography}

\end{document}